# CARBON DIOXIDE PRODUCTION RESPONSIBILITY ON THE BASIS OF COMPARING IN SITU AND MEAN CO$_2$ ATMOSPHERE CONCENTRATION DATA

S. Cht. Mavrodiev[1], L. Pekevski[2], B. Vachev[1]

[1]Institute for Nuclear Research and Nuclear Energy, Bulgarian academy of Sciences, Sofia, Bulgaria
[2]Skopje University, Skopje, Macedonia

*Abstract*

The method is proposed for estimation of regional CO$_2$ and other greenhouses and pollutants production responcibility. The comparison of CO2 local emissions reduction data with world CO2 atmosphere data will permit easy to judge for overall effect in curbing not only global warming but also chemical polution.

*Introduction*

One of the recent documents about climate change is the official report "Summary for Policymakers, Climate Change 2007: The Physical Science Basis" of United Nations Intergovernmental Panel for Climate Change (Summary for Policymakers, 2007) and Fourth Assessment Report (FAR) of the IPCC. The main conclusion is that with "90% confidence" humans have been the main drivers of warming since the 1950s. So, the climate anomalies as higher temperature, unusual variances of regional meteorological parameters as temperature, precipitations, increasing of the number and energy of hurricanes and the rising sea levels have anthropogenic origin.

Here is the time and place to remember one of the results of International scientific group "Clean and peaceful Black Sea" studying the development of the Black Sea ecology in the 80-th years of 20 century. Using the "Nuclear winter" lesson we tried to repeat it in the 1980'ties in the scale of the Black Sea region and to prevent the incoming ecological catastrophe by applying the "complex research" of the ecosystem by regularly publishing the "report for ecological conditions with scientific, business and management recommendation" for achievement the harmonic existence (stable development) of the region (Keodjian et al, 1990).

One of the results was that the Black Sea level is increasing with 1.5 +/-0.5 mm per year for the 1956- 1990 period as well as the evidence for increasing atmosphere exchange which leads to change of 15-20 Celsius degree themperature for 10- 15 hours . These facts could be interpreted as preliminary evidence of global warming (Mavrodiev, 1996, 1998)

Starting from the initial IPCC reports – First Assessment Report (FAR) (1990), Second Assessment Report (SAR) (1996) and Third Assessment Report (TAR) (2001) started discussions about reasons of climate change on the basis of critical analysis of the paleontology and today climate and weather data, balance and greenhouse gases cycle models, the variance of Sun conditions and Cosmic rays mass and energy spectra and so on. See for example Freitas (2002) and citations there.

The deep physical understanding of the Cosmic ray influence on the Climate via catalyzing the clouds formation was developed in the Climate balance model: Rusov at al. (2005). The important role of cosmic rays in clouds formation was investigated in the papers of Svensmark (2007a), Svensmark, Calder (2007b, 2007c).

In the last years in CERN started the accelerator study of the link between the cosmic rays and clouds – Kirkby (2001, 2007).

The cycles of atmosphere plankton started to be researched (see for example http://www.niwascience.co.nz), but till now there are not models and estimations for its influence on the atmosphere dynamics and climate parameters. The influence of organic aerosols on cloud properties and processes and on air quality and human health is one of the tasks of ACCENT EU project: http://www.accent-network.org/portal/joint-research-programme/aerosols.

Summarizing we could state that there are two possibilities:
- the global warming has anthropogenic origin,
- the visible change of today climate can be due to not well understood cosmic ray driven cloud aerosol generation and not well studied ocean stream periods, behavior and possible influence of plankton species distribution in ocean and atmosphere.

The model independent investigation of the hidden (heuristic) dependences and correlations between Space, Sun and Earth ecosystem parameters can help to estimate the reliability of different physical models as well as to give the basis for different time and space scales predictions ( Mavrodiev, Ries, 2006).

The aim of this paper is to illustrate a possible methodic for estimation of the regional $CO_2$ production responsibility using the simple comparison of model independent description of the $CO_2$ atmosphere annual mean concentration with in situ measurements.

*Experimental data*

The experimental methods for measuring in situ the $CO_2$ atmosphere concentration are published in ***http://cdiac.esd.ornl.gov/ trends/CO2/sio-keel.htm***
In the next Figure 1 are presented the GAW and SIO stations for measuring in situ the $CO_2$ atmosphere concentration.

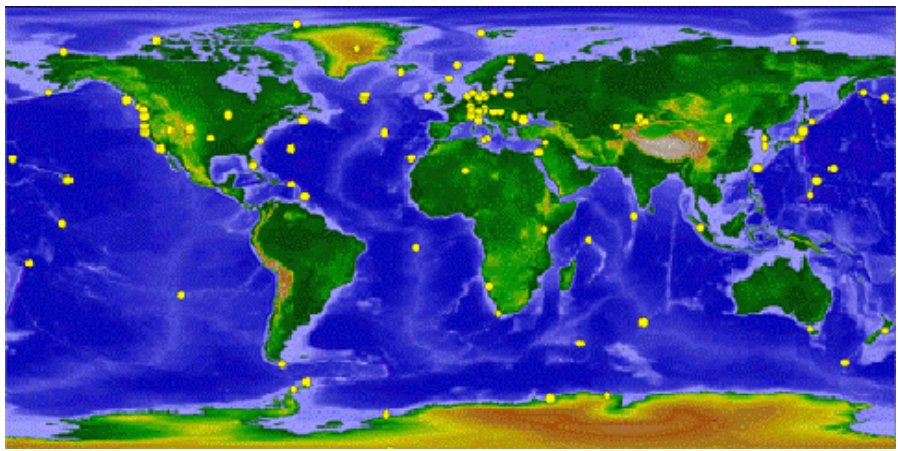

Figure 1. World map of the GAW and SIO in situ stations for measuring the $CO_2$ atmosphere concentration. The circles are the coordinates of 139 $CO_2$ stations. Data source: World Data Centre for Greenhouses Gases: http://gaw.kishou.go.jp/wdcgg/ Carbon Dioxide Research Group, SIO

*Estimation of regional $CO_2$ production responsibility*

The estimation is based on comparison of the annual mean $CO_2$ atmosphere concentration and regional in situ measured $CO_2$ concentration.

The description of annual mean $CO_2$ atmosphere concentration is present in the next figure. The explicit form of the fit model was discovered under the assumption (Mavrodiev, Ries, 2006) that the $CO_2$ atmosphere concentration depends on Sun spots number, Sun irradiation and $CO_2$ nature and anthropogenic production

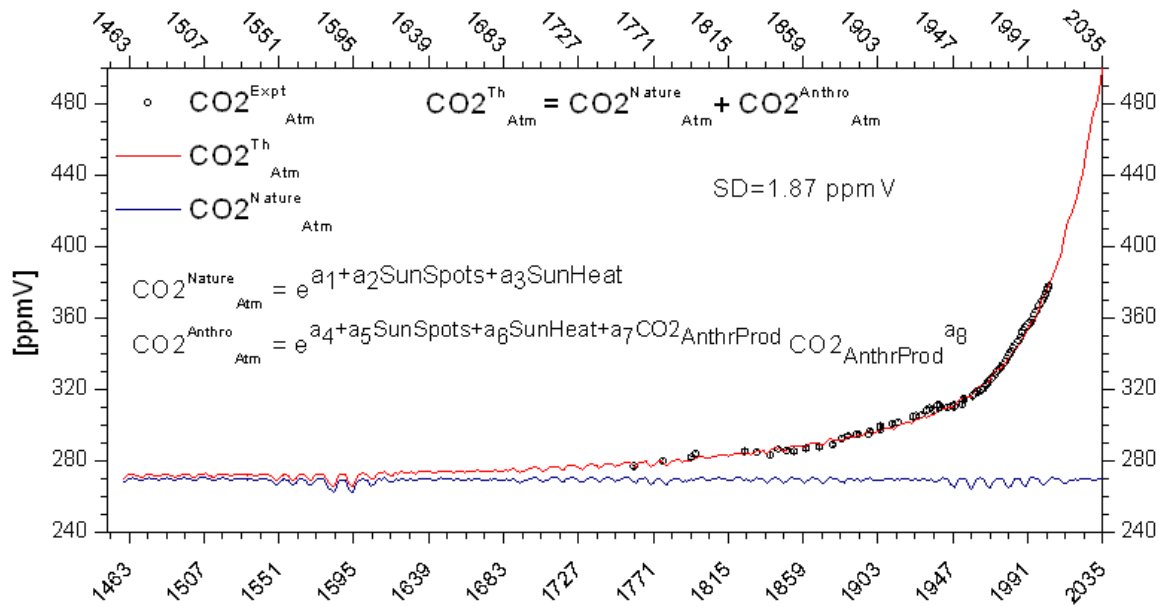

Figure 2. The $CO2^{Th}_{Atm}$ (yr) behavior as function of $CO2_{AnthroProduction}$, $Sun_{Spots}$ and $Sun_{Heat}$. <u>Data source</u> - Mauna Loa and Global $CO_2$ data: Atmospheric $CO_2$ concentrations (ppmv) derived from in situ air samples collected in Mauna Loa Observatory, Hawaii, C.D. Keeling, T.P. Whort, Carbon Dioxide Research Group, SIO; Global $CO_2$ Emissions from fossil-Fuel Burning, cement Manufacture and Gas *Flaming: 1751 – 2002, March, 10, 2005; Gregg Marland, Tom Boden, Robert J. Andres, Carbon Dioxide Information Analysis Center, Oak Ridge National Laboratory*

As one of the applications was mentioned the possibility for estimation of the regional $CO_2$ production responsibility by the value of linear fit parameter, different for stations.

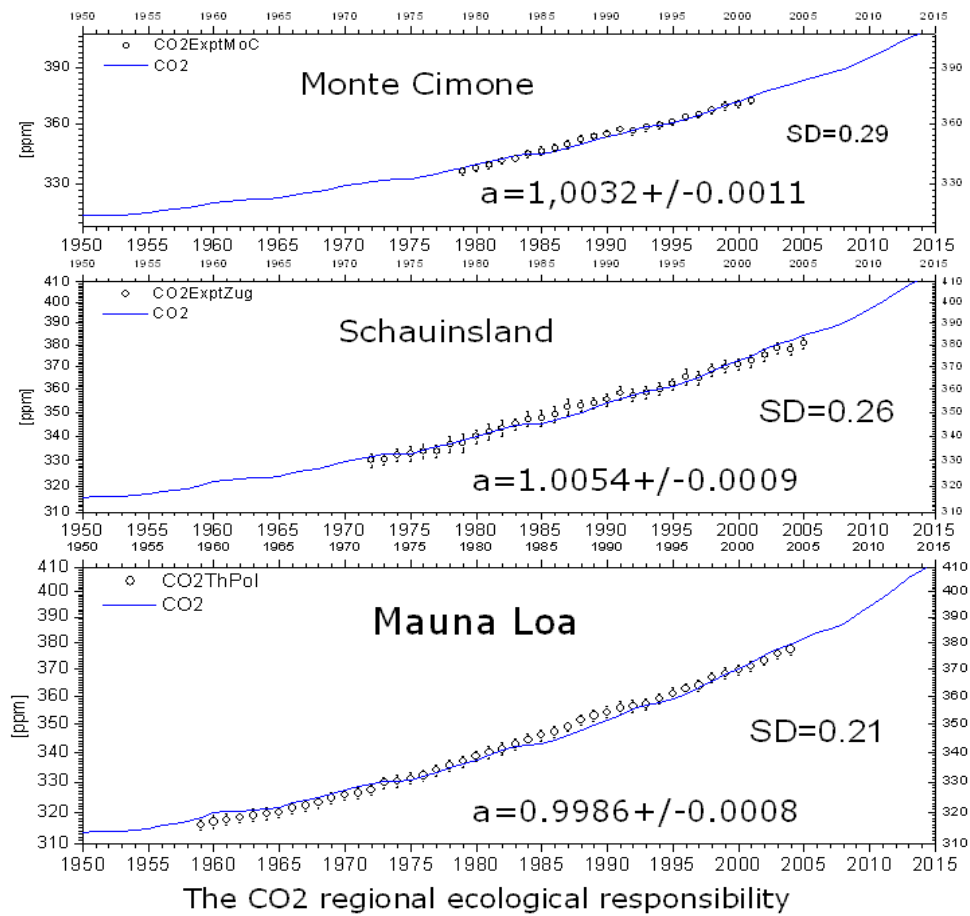

Figure 3. The possible estimation of regional $CO_2$ production responsibility for Monte Cimone, Schauinsland and Mauna Loa Stations. The blue curve is the mathematical model (fit) of $CO_2$ atmosphere concentration $CO2_{Atm}(yr)$, (Mavrodiev, Ries, 2006) Data source: World Data Centre for Greenhouses Gases
http://gaw.kishou.go.jp/wdcgg/

The explicit form of possible parametrization was generated and analysed using the Tichonov-Alexandrov inverse problem method and finally was recieved the next parametrization

**$CO2_{Local}(yr) = a(i_{local}) \exp(b*altitude) \, CO2_{Atm}(yr)$,**

where $a(i_{local})$, $i_{local}=1,\ldots,$Number of Station and b [km$^{-1}$] are the fit parameters.
**$CO2_{Local}(yr), CO2_{Atm}(yr)$**

The value of parameter $b = 6.7e^{-6}$ [m$^{-1}$] is illustrated in the next figure 4.

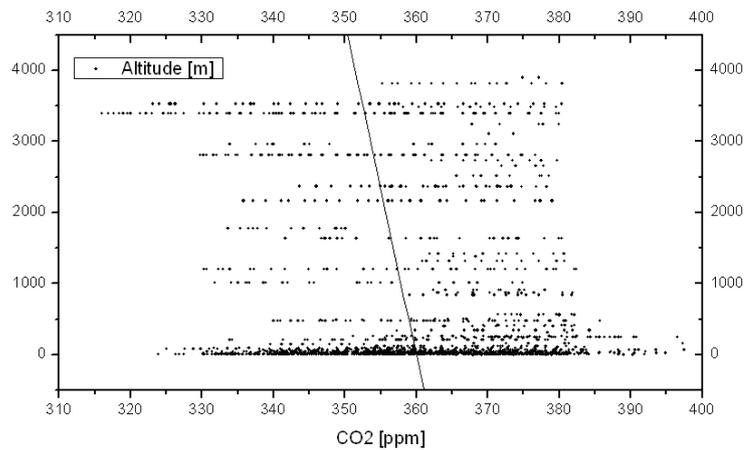

Figure 4. The altitude dependences exp(-b*Altitude) of $CO_2$ [ppm] atmosphere concentration, where b=6.71e$^{-6}$[m$^{-1}$], altitude [m] Data source: World Data Centre for Greenhouses Gases http://gaw.kishou.go.jp/wdcgg/; Carbon Dioxide Research Group, SIO, http://cdiac.esd.ornl.gov/trends/CO2/sio-keel.htm

The estimation of regional $CO_2$ production responsibility on the basis of in situ data $CO_2$ atmosphere concentration is presented on figure 5 and Table 1.

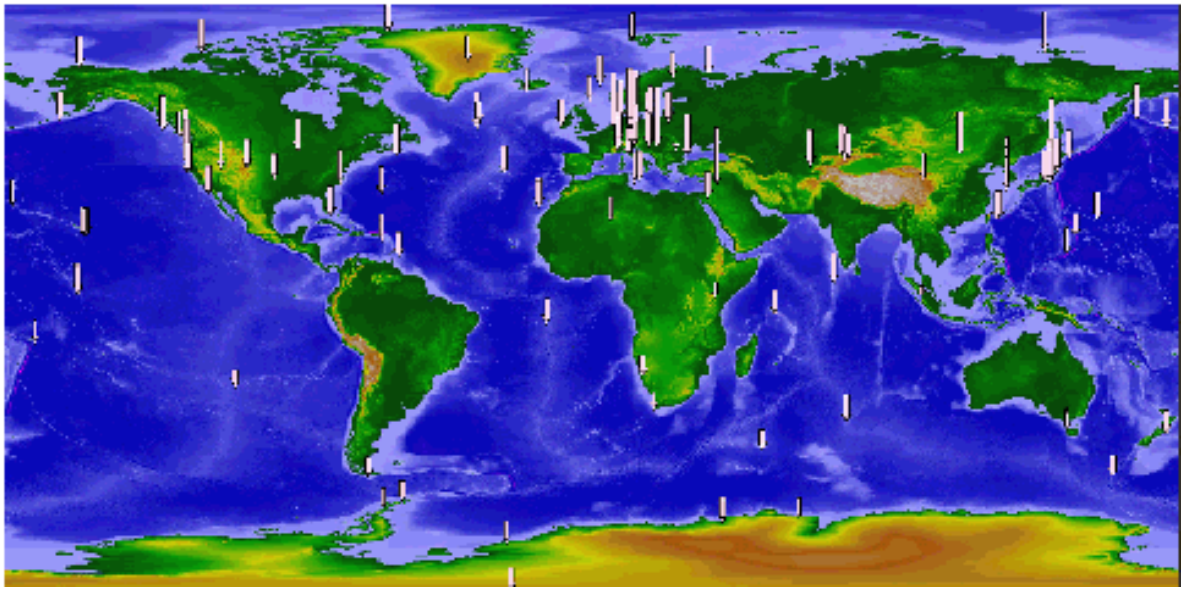

Figure 5. Estimation of regional $CO_2$ production responsibility on the basis of in situ data $CO_2$ atmosphere concentration. The circles are 139 $CO_2$ stations. The description of the annual CO2 data station is performing with function: CO2Station (yr) = a(iplaces) +exp(b*Altitude)*CO2Earth(yr), where a(iplaces) and b[1/km] are fit parameters and CO2Earth(yr) is the function describing the $CO_2$ atmosphere concentration. The bars are differences (a(iplaces) – 0.9700)*1000, which is relative estimation of the $CO_2$ regional production with mean $CO_2$ atmosphere concentration. Data source: World Data Centre for Greenhouses Gases http://gaw.kishow.go.jp/wdcgg.html; Carbon Dioxide Research Group, SIO, http://cdiac.esd.ornl.gov/trends/CO2/sio-keel.htm

Table 1. Estimation of regional $CO_2$ production responsibility on the basis of in situ data $CO_2$ atmosphere concentration

| Station | Hi2/NDF | CO2 production responsibility | Relative errror estimation | Latitude | Longitude | Altitude, m | Years |
|---|---|---|---|---|---|---|---|
| AmstIsl0 | 9.19E+00 | 2.46E+01 | 3.80E+01 | -3.78E+01 | 7.75E+01 | 65 | 25 |
| AmstIsl1 | 3.24E+00 | 3.07E+01 | 1.37E+01 | -3.80E+01 | 7.75E+01 | 150 | 9 |
| Ascension Island | 3.94E+00 | 2.79E+01 | 3.89E+01 | -7.92E+00 | -1.44E+01 | 54 | 26 |
| Assekrem | 2.74E-01 | 2.90E+01 | 5.70E+01 | 2.32E+01 | 5.42E+00 | 2728 | 10 |
| Cape Point | 6.19E+00 | 2.08E+01 | 2.24E+01 | -3.44E+01 | 1.85E+01 | 230 | 12 |
| Crozet | 5.30E+00 | 2.21E+01 | 2.51E+01 | -4.65E+01 | 5.19E+01 | 120 | 15 |
| Izana | 3.53E-01 | 3.33E+01 | 9.08E+01 | 2.83E+01 | -1.65E+01 | 2367 | 19 |
| Tenerife | 6.35E-01 | 2.98E+01 | 7.05E+01 | 2.83E+01 | -1.65E+01 | 2360 | 14 |
| Mt. Kenya | 2.26E-03 | 1.94E+01 | 1.56E+01 | -5.00E-01 | 3.73E+01 | 3897 | 2 |
| Gobabeb | 1.72E+00 | 2.32E+01 | 4.42E+00 | -2.36E+01 | 1.52E+01 | 461 | 2 |
| Mahe | 4.49E+00 | 2.79E+01 | 3.72E+01 | -4.67E+00 | 5.52E+01 | 7 | 26 |
| Anmyeon-do | 1.40E-01 | 5.20E+01 | 9.91E+00 | 3.65E+01 | 1.26E+02 | 47 | 6 |
| Bering Island | 3.74E-02 | 4.26E+01 | 9.76E+00 | 5.52E+01 | 1.66E+02 | 13 | 7 |
| Ochi-ishi | 5.38E-02 | 3.46E+01 | 1.10E+01 | 4.32E+01 | 1.46E+02 | 45 | 7 |
| Mt. Dodaira | 2.66E-01 | 4.14E+01 | 3.59E+01 | 3.60E+01 | 1.39E+02 | 840 | 13 |
| Hateruma | 2.66E-01 | 3.44E+01 | 1.35E+01 | 2.45E+01 | 1.24E+02 | 10 | 9 |
| Hamamatsu | 8.45E-02 | 7.43E+01 | 4.75E+00 | 3.47E+01 | 3.77E+01 | 35 | 3 |
| Issyk-Kul | 1.19E+00 | 4.21E+01 | 9.47E+01 | 4.26E+01 | 7.70E+01 | 1640 | 25 |
| Kaashidhoo | 7.24E-25 | 3.19E+01 | 1.49E+00 | 4.97E+00 | 7.35E+01 | 1 | 1 |
| Kisai | 4.60E-02 | 4.30E+01 | 6.39E+00 | 3.68E+01 | 1.40E+02 | 13 | 4 |
| Kotelny Island | 4.88E-02 | 4.73E+01 | 8.26E+00 | 7.60E+01 | 1.38E+02 | 5 | 6 |
| Gosan | 1.31E-01 | 3.87E+01 | 1.89E+01 | 3.33E+01 | 1.26E+02 | 72 | 12 |
| Kyzylcha | 1.04E-22 | 4.54E+01 | 1.85E+00 | 4.09E+01 | 6.62E+01 | 340 | 1 |
| Sary Taukum | 4.57E-02 | 3.48E+01 | 1.76E+01 | 4.45E+01 | 7.56E+01 | 412 | 8 |
| Plateau Assy | 5.53E-02 | 2.72E+01 | 4.36E+01 | 4.33E+01 | 7.79E+01 | 2519 | 8 |
| Minamitorishima | 3.68E-01 | 3.03E+01 | 1.98E+01 | 2.43E+01 | 1.54E+02 | 8 | 13 |
| Ryori | 1.75E-01 | 3.80E+01 | 3.52E+01 | 3.93E+01 | 1.42E+02 | 260 | 19 |
| Tae-ahn Peninsula | 1.06E-01 | 3.92E+01 | 2.29E+01 | 3.67E+01 | 1.26E+02 | 20 | 15 |
| Tsukuba | 3.86E-02 | 1.03E+02 | 1.89E+01 | 3.65E+01 | 1.40E+02 | 26 | 13 |
| Takayama | 1.55E-01 | 3.75E+01 | 4.03E+01 | 3.61E+01 | 1.37E+02 | 1420 | 11 |
| Urawa | 1.51E-01 | 5.27E+01 | 1.33E+01 | 3.59E+01 | 1.40E+02 | 10 | 9 |
| Ulaan | 1.51E-01 | 5.33E+01 | 2.54E+01 | 4.45E+01 | 1.12E+02 | 914 | 9 |
| Waliguan | 4.80E-02 | 3.35E+01 | 2.11E+01 | 3.63E+01 | 1.01E+02 | 3810 | 3 |
| Mt. Waliguan | 1.60E-01 | 3.12E+01 | 9.37E+01 | 3.63E+01 | 1.01E+02 | 3810 | 13 |
| Yonagunijima | 1.98E-01 | 3.27E+01 | 1.43E+01 | 2.45E+01 | 1.23E+02 | 30 | 9 |
| Easter Island | 3.00E+00 | 1.88E+01 | 1.76E+01 | -2.71E+01 | -1.09E+02 | 50 | 11 |
| Tierra del Fuego | 2.06E+00 | 2.03E+01 | 9.35E+00 | -5.49E+01 | -6.85E+01 | 20 | 6 |
| Alert | 4.20E-02 | 3.85E+01 | 1.75E+01 | 8.25E+01 | -6.25E+01 | 210 | 11 |
| Alert | 1.64E-01 | 3.44E+01 | 3.34E+01 | 8.25E+01 | -6.25E+01 | 210 | 19 |
| Alert | 7.99E-02 | 3.13E+01 | 2.55E+01 | 8.25E+01 | -6.25E+01 | 210 | 14 |
| St. Croix | 3.37E-01 | 3.40E+01 | 1.32E+01 | 1.78E+01 | -6.48E+01 | 3 | 10 |
| St. David's Head | 3.94E-01 | 3.17E+01 | 2.59E+01 | 3.24E+01 | -6.47E+01 | 30 | 17 |
| Tudor Hill | 3.37E-01 | 3.40E+01 | 1.35E+01 | 3.23E+01 | -6.49E+01 | 30 | 10 |
| Barrow | 8.21E-02 | 3.55E+01 | 4.15E+01 | 7.13E+01 | -1.57E+02 | 8 | 30 |
| Barrow | 9.23E-02 | 3.48E+01 | 4.73E+01 | 7.13E+01 | -1.57E+02 | 11 | 34 |
| Cold Ba | 1.25E-01 | 3.44E+01 | 3.30E+01 | 5.52E+01 | -1.63E+02 | 25 | 23 |
| Cape Meares | 1.95E-01 | 3.73E+01 | 2.11E+01 | 4.55E+01 | -1.24E+02 | 30 | 15 |
| Cape St. James | 1.15E-02 | 4.08E+01 | 5.97E+00 | 5.19E+01 | -1.31E+02 | 89 | 4 |
| Estevan Point | 3.21E-02 | 3.24E+01 | 1.23E+01 | 4.94E+01 | -1.27E+02 | 39 | 8 |
| Grifton | 4.91E-02 | 4.20E+01 | 1.31E+01 | 3.54E+01 | -7.74E+01 | 505 | 6 |
| Key Biscayne | 5.56E-01 | 3.45E+01 | 4.20E+01 | 2.57E+01 | -8.02E+01 | 3 | 30 |
| Park Falls | 2.41E-02 | 3.60E+01 | 2.84E+01 | 4.59E+01 | -9.03E+01 | 868 | 10 |
| Mould Bay | 6.64E-02 | 3.84E+01 | 2.30E+01 | 7.63E+01 | -1.19E+02 | 58 | 16 |
| Niwot Ridge | 4.63E-01 | 3.30E+01 | 2.22E+02 | 4.03E+01 | -1.06E+02 | 3526 | 36 |
| Olympic Peninsula | 4.66E-02 | 4.11E+01 | 9.96E+00 | 4.83E+01 | -1.24E+02 | 488 | 5 |
| Point Arena | 7.01E-02 | 3.42E+01 | 8.03E+00 | 3.90E+01 | -1.24E+02 | 17 | 5 |
| Ragged Point | 7.56E-01 | 2.97E+01 | 2.77E+01 | 1.32E+01 | -5.94E+01 | 45 | 18 |
| Sable Island | 9.48E-02 | 3.87E+01 | 1.45E+01 | 4.39E+01 | -6.02E+01 | 5 | 11 |
| Southern Great Plains | 3.97E-02 | 2.82E+01 | 6.32E+00 | 3.68E+01 | -9.75E+01 | 314 | 3 |
| Shemya Island | 1.13E-01 | 3.44E+01 | 2.87E+01 | 5.27E+01 | 1.75E+02 | 40 | 19 |
| Trinidad Head | 7.67E-03 | 3.49E+01 | 5.33E+00 | 4.13E+01 | -1.24E+02 | 107 | 3 |
| Wendover | 2.55E-01 | 3.21E+01 | 4.23E+01 | 3.99E+01 | -1.14E+02 | 1320 | 12 |
| Baring Head | 3.47E+00 | 2.64E+01 | 3.59E+01 | -4.14E+01 | 1.75E+02 | 85 | 25 |
| Bukit Koto Tabang | 5.95E-03 | 1.16E+01 | 5.96E+00 | 2.00E-01 | 1.00E+02 | 865 | 2 |
| Cape Ferguson | 2.89E+00 | 2.36E+01 | 2.11E+01 | 1.93E+01 | 1.48E+02 | 2 | 14 |
| Cape Grim | 5.88E+00 | 2.38E+01 | 3.33E+01 | -4.07E+01 | 1.45E+02 | 94 | 21 |
| Cape Grim | 3.04E+00 | 2.12E+01 | 2.30E+01 | -4.07E+01 | 1.45E+02 | 94 | 14 |

| Station | | | | | | | |
|---|---|---|---|---|---|---|---|
| Cape Grim | 1.25E-01 | 2.53E+01 | 5.48E+00 | -4.07E+01 | 1.45E+02 | 94 | 4 |
| Cape Grim | 2.10E-02 | 2.58E+01 | 4.13E+00 | -4.07E+01 | 1.45E+02 | 94 | 3 |
| Christmas Island | 8.23E+00 | 3.15E+01 | 2.31E+01 | 1.70E+00 | -1.57E+02 | 3 | 16 |
| Guam | 1.08E+00 | 3.18E+01 | 3.83E+01 | 1.34E+01 | 1.45E+02 | 2 | 27 |
| Cape Kumukahi | 3.74E-01 | 3.16E+01 | 4.08E+01 | 1.95E+01 | -1.55E+02 | 3 | 29 |
| Sand Island | 7.39E-01 | 3.24E+01 | 2.94E+01 | 2.82E+01 | -1.77E+02 | 8 | 20 |
| Mauna Loa | 6.38E-01 | 3.35E+01 | 1.82E+02 | 1.95E+01 | -1.56E+02 | 3397 | 30 |
| Mauna Loa | 7.05E-01 | 3.30E+01 | 1.84E+02 | 1.95E+01 | -1.56E+02 | 3397 | 30 |
| Mauna Loa | 5.43E-01 | 3.04E+01 | 9.22E+01 | 1.95E+01 | -1.56E+02 | 3397 | 14 |
| Macquarie Island | 2.96E+00 | 2.06E+01 | 2.13E+01 | -5.45E+01 | 1.59E+02 | 12 | 14 |
| Tutuila | 1.31E+01 | 2.77E+01 | 4.07E+01 | -1.42E+01 | -1.71E+02 | 42 | 28 |
| Tutuila | 1.08E+01 | 2.75E+01 | 4.62E+01 | -1.42E+01 | -1.71E+02 | 42 | 32 |
| Terceira Island | 4.71E-01 | 3.14E+01 | 2.95E+01 | 3.88E+01 | -2.74E+01 | 40 | 20 |
| Baltic Sea | 5.82E-02 | 4.07E+01 | 1.69E+01 | 5.54E+01 | 1.72E+01 | 28 | 11 |
| Brotjacklriegel | 9.90E-02 | 4.31E+01 | 6.74E+01 | 4.88E+01 | 1.32E+01 | 1016 | 25 |
| Black Sea | 9.49E-02 | 4.89E+01 | 1.69E+01 | 4.42E+01 | 2.87E+01 | 3 | 11 |
| Monte Cimone | 2.18E-01 | 3.29E+01 | 1.21E+02 | 4.42E+01 | 1.07E+01 | 2165 | 27 |
| Deuselbach | 3.21E-01 | 5.59E+01 | 5.01E+01 | 4.98E+01 | 7.50E+00 | 480 | 25 |
| Puszcza Borec./Diabla Gora | 9.25E-01 | 3.71E+01 | 1.09E+01 | 5.42E+01 | 2.27E+01 | 157 | 6 |
| Fundata | 1.66E-01 | 2.31E+00 | 7.42E+00 | 4.55E+01 | 2.53E+01 | 1384 | 2 |
| Dwejra Point | 2.99E-02 | 3.48E+01 | 4.44E+00 | 3.65E+01 | 1.42E+01 | 30 | 3 |
| Hegyhatsal | 2.56E-01 | 6.89E+01 | 1.54E+01 | 4.70E+01 | 1.67E+01 | 248 | 8 |
| Hegyhatsal | 1.08E-01 | 5.35E+01 | 2.11E+01 | 4.70E+01 | 1.67E+01 | 248 | 11 |
| Hegyhatsal | 7.29E-02 | 4.85E+01 | 1.91E+01 | 4.70E+01 | 1.67E+01 | 248 | 10 |
| Hegyhatsal | 1.00E-01 | 4.43E+01 | 2.10E+01 | 4.70E+01 | 1.67E+01 | 248 | 11 |
| Hegyhatsal | 2.95E-02 | 4.05E+01 | 2.65E+01 | 4.70E+01 | 1.66E+01 | 344 | 13 |
| Heimaey | 7.07E-02 | 2.91E+01 | 2.16E+01 | 6.33E+01 | -2.03E+01 | 100 | 13 |
| Kollumerwaard | 6.65E-01 | 6.40E+01 | 1.32E+01 | 5.33E+01 | 6.28E+00 | 0 | 9 |
| K-puszta | 2.31E-01 | 7.54E+01 | 2.35E+01 | 4.70E+01 | 1.96E+01 | 125 | 15 |
| Waldhof | 2.76E-01 | 7.38E+01 | 4.52E+01 | 5.28E+01 | 1.08E+01 | 74 | 31 |
| Lampedusa | 1.40E-01 | 3.50E+01 | 2.05E+01 | 3.55E+01 | 1.26E+01 | 45 | 13 |
| Mace Head | 8.84E-02 | 2.89E+01 | 2.16E+01 | 5.33E+01 | -9.90E+00 | 25 | 14 |
| Mace Head | 1.16E-01 | 2.82E+01 | 1.85E+01 | 5.32E+01 | -9.90E+00 | 25 | 12 |
| Neuglobsow | 1.96E-01 | 7.17E+01 | 1.78E+01 | 5.32E+01 | 1.33E+01 | 65 | 11 |
| Pallas-Sammaltunturi | 5.59E-02 | 3.14E+01 | 1.72E+01 | 6.80E+01 | 2.41E+01 | 565 | 7 |
| Pallas-Sammaltunturi | 6.30E-03 | 2.91E+01 | 9.95E+00 | 6.80E+01 | 2.41E+01 | 560 | 4 |
| Plateau Rosa | 1.35E-01 | 2.96E+01 | 8.13E+01 | 4.59E+01 | 7.70E+00 | 3480 | 12 |
| Zugspitze/Schneefernerhaus | 5.47E+00 | 2.15E+01 | 2.31E+01 | 4.74E+01 | 1.10E+01 | 2656 | 4 |
| Shetland | 8.53E-02 | 2.84E+01 | 1.69E+01 | 6.08E+01 | -1.25E+00 | 30 | 11 |
| Sonnblick | 4.45E-03 | 2.56E+01 | 1.27E+01 | 4.75E+01 | 1.30E+01 | 3106 | 2 |
| Schauinsland | 9.83E-02 | 3.64E+01 | 1.00E+02 | 4.79E+01 | 7.92E+00 | 1205 | 33 |
| Schauinsland | 3.83E-03 | 3.01E+01 | 2.08E+01 | 4.79E+01 | 7.92E+00 | 1205 | 6 |
| Ocean Station Charlie | 6.38E-03 | 4.27E+01 | 5.39E+00 | 5.28E+01 | -3.55E+01 | 5 | 4 |
| Ocean Stat C | 9.95E-03 | 2.35E+01 | 4.82E+01 | 5.40E+01 | -3.50E+01 | 6 | 4 |
| Ocean Stat 'M' | 1.32E-01 | 3.27E+01 | 3.59E+01 | 6.60E+01 | 2.00E+00 | 5 | 25 |
| Summit | 1.58E-01 | 3.05E+01 | 3.96E+01 | 7.26E+01 | -3.85E+01 | 3238 | 6 |
| Teriberka | 8.42E-02 | 3.39E+01 | 2.31E+01 | 6.92E+01 | 3.51E+01 | 40 | 15 |
| Sede Boker | 1.12E-01 | 3.14E+01 | 2.15E+01 | 3.11E+01 | 3.49E+01 | 400 | 10 |
| Wank Peak | 5.60E-02 | 1.36E+01 | 4.81E+01 | 4.75E+01 | 1.12E+01 | 1780 | 13 |
| Westerland | 1.46E-01 | 4.90E+01 | 3.84E+01 | 5.49E+01 | 8.32E+00 | 12 | 28 |
| Zeppelinfjellet | 2.46E-01 | 3.30E+01 | 3.03E+01 | 7.89E+01 | 1.19E+01 | 475 | 14 |
| Ny-Alesund | 6.26E-02 | 3.12E+01 | 2.69E+01 | 7.89E+01 | 1.19E+01 | 475 | 12 |
| Zugspitze | 4.76E-02 | 3.09E+01 | 4.11E+01 | 4.74E+01 | 1.10E+01 | 2960 | 7 |
| Zingst | 6.38E-02 | 5.33E+01 | 7.64E+00 | 5.44E+01 | 1.27E+01 | 1 | 5 |
| Zugspitze | 7.51E-02 | 1.81E+01 | 4.65E+01 | 4.74E+01 | 1.10E+01 | 2960 | 9 |
| Halley Bay | 5.44E+00 | 2.42E+01 | 3.14E+01 | -7.56E+01 | -2.65E+01 | 33 | 21 |
| Jubany | 3.56E+00 | 2.13E+01 | 1.69E+01 | -6.22E+01 | -5.87E+01 | 15 | 11 |
| Mawson | 3.28E+00 | 2.14E+01 | 2.32E+01 | -6.76E+01 | 6.29E+01 | 32 | 15 |
| Palmer Station | 3.14E+00 | 2.47E+01 | 3.72E+01 | -6.49E+01 | -6.40E+01 | 10 | 26 |
| South Pole | 3.68E+00 | 2.68E+01 | 1.48E+02 | -9.00E+01 | -2.48E+01 | 2810 | 28 |
| South Pole | 3.74E+00 | 2.70E+01 | 1.53E+02 | -9.00E+01 | -2.48E+01 | 2810 | 29 |
| South Pole | 3.84E+00 | 2.65E+01 | 1.43E+02 | -9.00E+01 | -2.48E+01 | 2810 | 27 |
| Syowa Station | 2.77E-01 | 3.08E+01 | 6.81E+00 | -6.90E+01 | 3.96E+01 | 29 | 5 |
| Syowa Station | 6.53E+00 | 2.35E+01 | 2.53E+01 | -6.90E+01 | 3.96E+01 | 14 | 17 |
| Alert, NWT | 1.58E-01 | 3.48E+01 | 3.34E+01 | 8.25E+01 | -6.25E+01 | 210 | 19 |
| Barrow | 8.53E-02 | 3.53E+01 | 4.19E+01 | 7.13E+01 | -1.57E+02 | 11 | 30 |
| La Jolla Pier | 2.52E-01 | 3.27E+01 | 3.68E+01 | 3.22E+01 | -1.18E+02 | 10 | 26 |
| Mauna Loa | 9.41E-01 | 3.11E+01 | 2.64E+01 | 1.95E+01 | -1.56E+02 | 3397 | 46 |
| Cape Kumukahi | 4.25E-01 | 3.24E+01 | 3.54E+01 | 1.95E+01 | -1.55E+02 | 3 | 25 |
| Christmas Island | 1.98E+00 | 3.26E+01 | 3.10E+01 | 2.00E+00 | -1.57E+02 | 3 | 23 |
| Cape Matatula | 1.05E+01 | 2.83E+01 | 3.38E+01 | -1.43E+01 | -1.71E+02 | 30 | 23 |
| South Pole | 3.82E+00 | 2.70E+01 | 1.53E+02 | -9.00E+01 | -2.48E+01 | 2810 | 29 |

*Illustration of the methodology and interpretation*

In the next figure are presented comparison the CO2 annual mean atmosphere in situ data and the model independent description of annual atmosphere mean concentration.

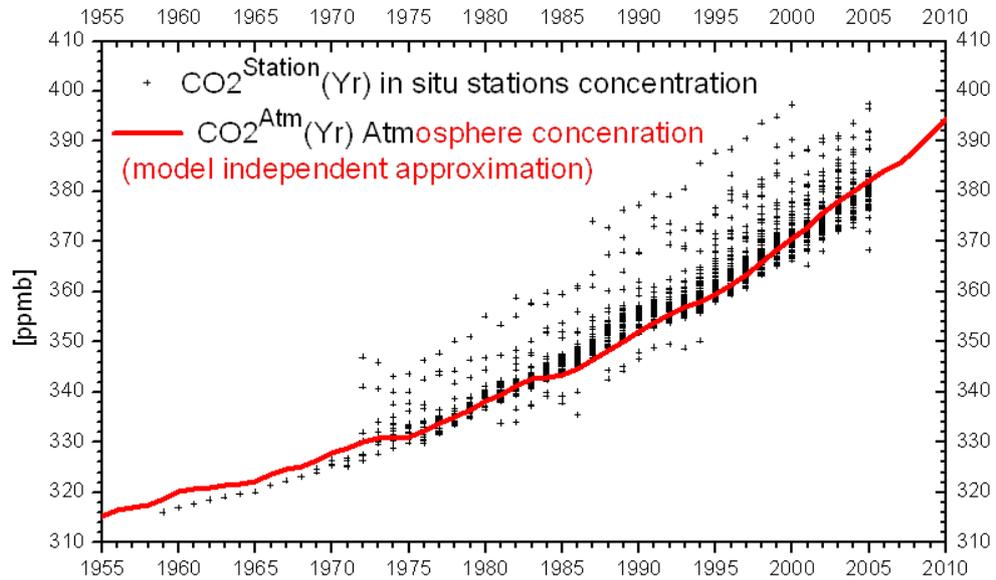

Figure 6. Comparison of model independent CO2 annual mean atmosphere concentration with in situ data (139 stations)

The linear fit estimation of trend of world data for regioanal CO2 production responcibility is presented in the next figure.

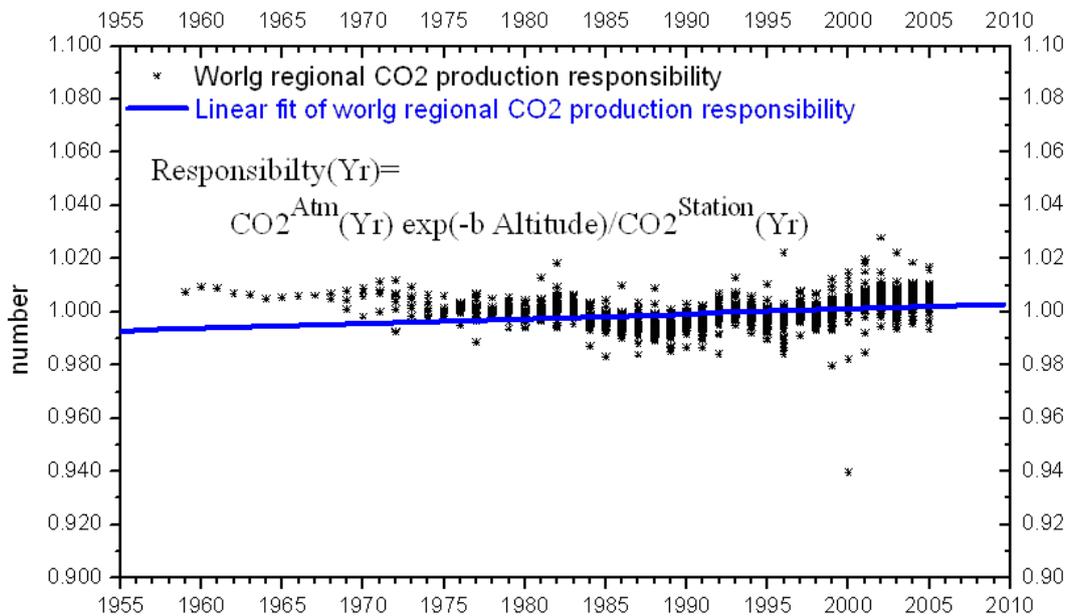

Figure 7. Liner fit of all data for CO2 productiomn responcibility

**LinFit= (0.63 +/- 0.03) + (1.84 +/- 0.13)*10$^{-4}$Yr**

In the next 3 figure (Figure 8,9,10) are presented the temperature behavior and the behavior of function Responsibility for stations Monte Cimone, Schauinsland and Waldhof.

Figure 8, 9, 10. Monte Cimone, Schauinsland and Waldhof temperature and $CO_2$ data

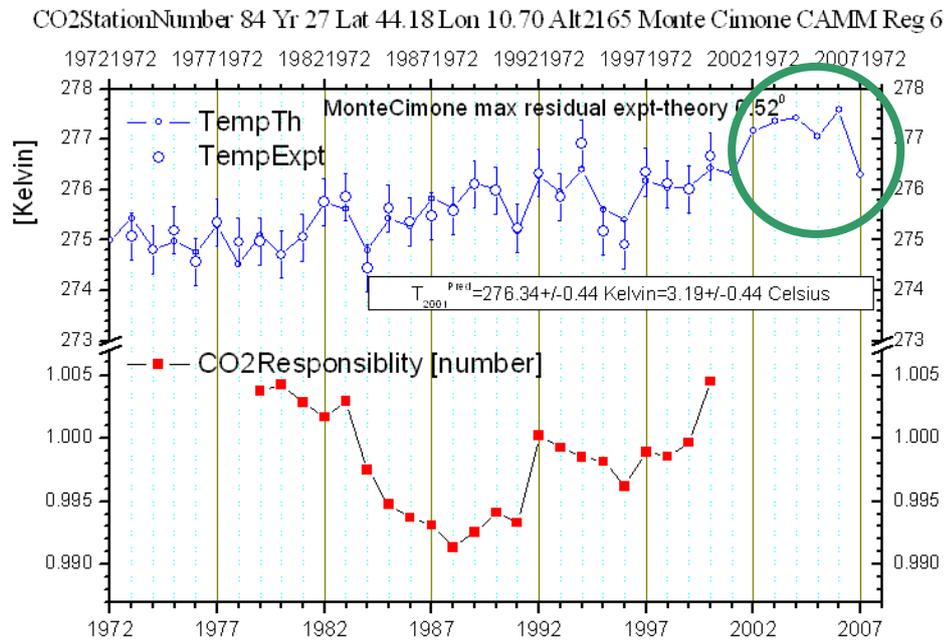

The visible negative and positive correlations between local annual mean temperature and CO2 regional production responsibility

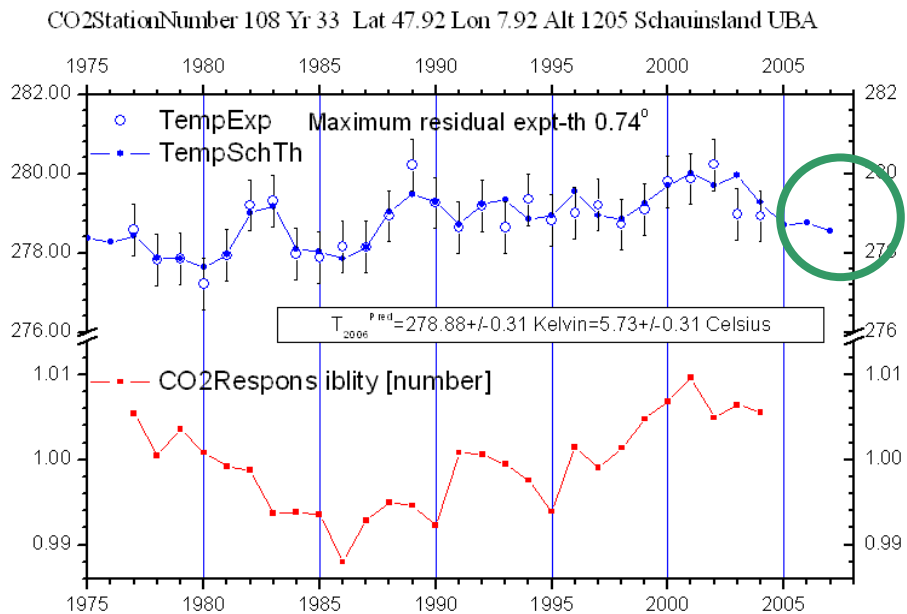

The visible negative correlation between local annual mean temperature and CO2 regional production responsibility

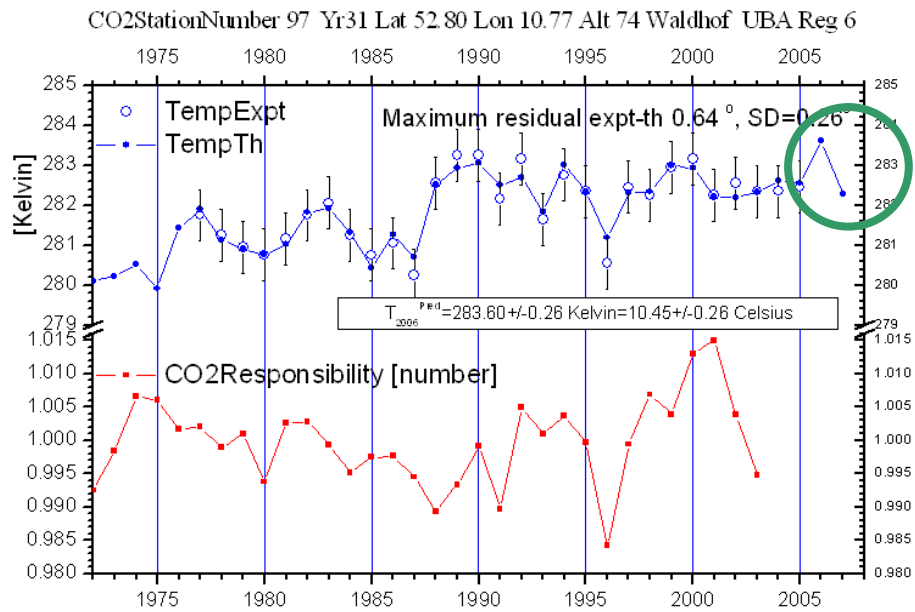

The complicated Waldhop negative and positive correlations between local annual mean temperature and CO2 regional production responsibility

    The correlation between temperature and $CO_2$ regional production responsibility is more complicated for the Monte Cimone and Waldhof because of the influence of Mediterranean and North see. The interpretation have to include;
- the physics of Climate model, where the increasing of the $CO_2$ atmosphere concentration is mainly concequences of temperature increasing
- the meteorological data for atmosphere exchange as well as the data for energy production and consummation as well as for car traffic.

    But for Schauinsland there is seen clear negative correlation between higher temperature and less $CO_2$ production in the next time period. The estimation of the inertion of such correlation can serve as estimation of the effective planning and management of the regional energy resources and energy consumption.

*Conclusion*
    The described methical example can serve as initial bases for estimation of regional $CO_2$ and other greenhouses and pollutants production. The including of the analysis the data for industry, agriculture, transport and cities infrastructure can permit to create not only a model for regional pollution responsibility but also a model for helping the decision makers planning.
    The understanding of the connection between parameters of the regional mathematical temperature model and middle weather prediction can permit to plan more accuracy the next year energy resources.

*Acknowledgments*

The authors are grateful to Cht. S Mavrodiev and I. Kalapov, BEO INRNE, BAS for the team work, to Dr. Wolfgang Fricke, Hohenpeissenberg Meteorological Observatory, Germany Meteorological Service and Dr. Ludwig Ries, Germany Environmental Agency for critical discussions of the use of results of inverse problems method and for the Hohenpeissenberg century temperature data, to Dr. Paolo Bonasoni and Dr. P. Cristofanelli, ISAC, CNR, Bologna, Italy for information support. The partly support of this work by UFS Zugspitze and UBA, Germany and the BEOBAL FP6 project (INCO-CT-2005-016663) is highly appreciated.